# Positioning for the Internet of Things:
# A 3GPP Perspective

Xingqin Lin, Johan Bergman, Fredrik Gunnarsson, Olof Liberg, Sara Modarres Razavi,

Hazhir Shokri Razaghi, Henrik Rydén, and Yutao Sui

Ericsson

Contact: xingqin.lin@ericsson.com

*Abstract—* **Many use cases in the Internet of Things (IoT) will require or benefit from location information, making positioning a vital dimension of the IoT. The 3rd Generation Partnership Project (3GPP) has dedicated a significant effort during its Release 14 to enhance positioning support for its IoT technologies to further improve the 3GPP-based IoT eco-system. In this article, we identify the design challenges of positioning support in Long-Term Evolution Machine Type Communication (LTE-M) and Narrowband IoT (NB-IoT), and overview the 3GPP's work in enhancing the positioning support for LTE-M and NB-IoT. We focus on Observed Time Difference of Arrival (OTDOA), which is a downlink based positioning method. We provide an overview of the OTDOA architecture and protocols, summarize the designs of OTDOA positioning reference signals, and present simulation results to illustrate the positioning performance.**

## I. INTRODUCTION

In this article, we provide an overview of the positioning techniques developed by the 3rd Generation Partnership Project (3GPP) for location based services (LBS) in massive Internet of Things (IoT) devices. The IoT is a vision for the future world where everything that can benefit from a connection will be connected. Cellular technologies are being developed or evolved to play an indispensable role in the IoT world, particularly the machine type communication (MTC) [1]. MTC is characterized by lower demands on data rates than mobile broadband, but with higher requirements on e.g. low cost device design, better coverage, and the ability to operate for years on batteries without charging or replacing the batteries [2]. Prospective applications include utility metering, environment monitoring, asset tracking, and municipal light and waste management. Providing connectivity for such use cases with massive number of devices is also considered to be part of the requirements for next generation mobile telecommunications (a.k.a. 5G) [3].

The 3GPP dedicated a major effort during its Release 13 to develop cellular systems that provide low power wide area IoT connectivity. Two prominent 3GPP MTC technologies are the Long-Term Evolution (LTE) MTC (LTE-M) and Narrowband IoT (NB-IoT). With a minimum system bandwidth of 1.4 MHz, LTE-M is based on LTE and incorporates additional improvements to better support IoT services [4]. NB-IoT is a new radio access technology that uses a system bandwidth of 180 kHz to provide more flexible deployment options [5]. Both technologies promise to provide improved coverage for massive number of low-throughput low-cost devices with low device power consumption in delay-tolerant applications

The market for low power wide area IoT connectivity shows high demand from operators and governments. It is expected that a significant portion of IoT applications will need LBS, making positioning a vital dimension of the IoT. Example IoT use cases requiring positioning services include logistics tracking, wearables, and animal husbandry and fishery [6][7][8]. The widely used global navigation satellite system (GNSS) positioning method is not suitable for massive IoT, due to affordability issues in terms of power and cost to support GNSS chip. Besides, the IoT devices may be in extreme coverage challenging locations (e.g. indoors, encapsulated, or underground) that are not covered by GNSS. A promising alternative is to position based on terrestrial cellular networks.

In Release 13, only limited positioning support was available for LTE-M and NB-IoT. Hence, enhancing the 3GPP positioning methods is necessary to improve the 3GPP-based IoT eco-system. This motivated 3GPP to dedicate a major effort in Release 14 to improve the positioning support for both LTE-M and NB-IoT, with an ambition to further increase the market impact of 3GPP MTC technologies.

## II. USE CASES AND CHALLENGES OF IOT POSITIONING

Positioning has been supported for LTE since Release 9 and mainly to meet regulatory emergency call positioning requirements. There have been already inputs from network operators on the positioning requirements and potential use cases of massive IoT positioning [6][7][8]. For example, it was estimated that 75% of the evaluated use cases would require or at least benefit from positioning information [8]. Some example use cases include *smart "things"* (wearable, machinery control, safety monitoring, smart bicycles, parking sensors, etc.), *transport "things"* (asset tracking, pet tracking, livestock tracking, etc.), and *sensing "things"* (environment monitoring, smoke detectors, gas/water/electricity metering, etc.).

Different use cases have different positioning accuracy requirements, which may range from a few meters to hundreds of meters. Though not explicitly specified, achieving 50 m horizontal positioning accuracy has been the benchmark design target for the positioning support in LTE-M and NB-IoT.

3GPP has focused on OTDOA (Observed Time Difference of Arrival) for both LTE-M and NB-IoT positioning in Release 14, after a careful consideration of several important factors including positioning accuracy, user equipment (UE) complexity impact, UE power consumption, and standardization and network impacts. To improve OTDOA support for LTE-M and NB-IoT, the enhancements need to address two key design challenges.

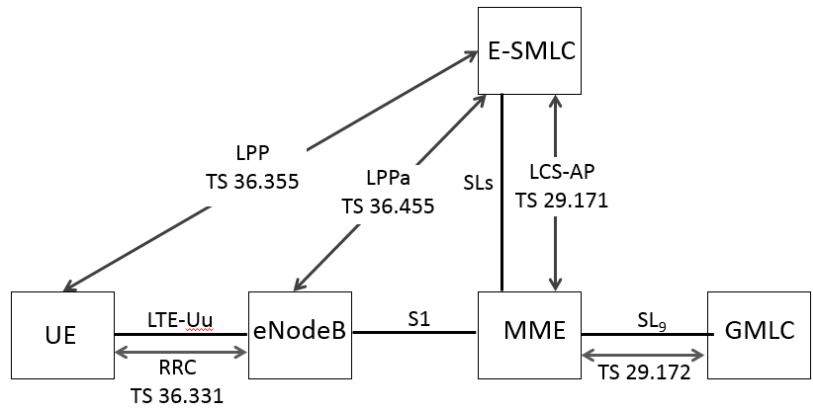

Figure 1: LTE positioning architecture

- *Limited device capability:* LTE-M and NB-IoT UEs are of low lost and limited capability. The UE only has a single receive RF chain. The maximum channel bandwidths of low complexity bandwidth reduced LTE-M and NB-IoT UEs are 1.4 MHz and 180 kHz, respectively, though the more advanced LTE-M UE in Release 14 can receive 5 MHz wide signal. In contrast, normal LTE UE can receive 20 MHz wide signal (and larger with carrier aggregation). The ranging/timing resolution is fundamentally limited by the signal bandwidth. It is challenging to provide satisfactory positioning support for the low complexity bandwidth reduced devices.

- *Coverage extension:* LTE-M and NB-IoT UEs may be in extreme coverage challenging locations. LTE-M has been designed to enhance the coverage with 15 dB compared to normal LTE coverage, while NB-IoT has been designed to enhance the coverage with 20 dB compared to extreme GSM coverage. For devices located in extreme coverage challenging locations, the signal to noise ratios are low, usually below -10 dB. How to improve OTDOA positioning support for the extreme coverage challenging locations is another key design challenge.

## III. PRELIMINARIES OF OTDOA ARCHITECTURE AND PROTOCOLS

In this section, we present the preliminaries of OTDOA architecture and protocols developed by 3GPP. Interested readers may refer to e.g. [10] and the related 3GPP technical specifications for more details [9][11].

In OTDOA, UE measures the time of arrival (TOA) of positioning reference signals (PRS) received from multiple cells, and subtracts the TOA of a reference cell from the measured TOAs to form the reference signal time difference (RSTD) measurements, which are the time different of arrival (TDOA). Geometrically, each TDOA constrains the desired UE's position to a hyperbola. If the TOA measurements were noise and interference free, the hyperbolas would intersect at one point that is the desired UE position.

Positioning in LTE is supported by the architecture illustrated in Figure 1. A typical positioning signaling flow under the architecture may go as follows. First, the Mobility Management Entity (MME) initiates a location service or receives a location service request from the UE or Gateway Mobile Location Centre (GMLC). GMLC is the first node with which an external LBS client communicates. Second, the MME sends a positioning request to the location server, Evolved Serving Mobile Location Centre (E-SMLC). The E-SMLC processes the request, communicates with the UE, and requests for RSTD measurements. Upon receiving RSTD measurements from the UE, the E-SMLC estimates the UE's position and sends the result back to the MME. The MME may further forward the result to the UE or GMLC as appropriate.

The signaling between the E-SMLC and UE is carried out via the LTE Positioning Protocol (LPP). Moreover, there are also interactions between the E-SMLC and Evolved Node B (eNB) via the LPP A (LPPa) protocol, to some extent supported by the interactions between the eNB and UE via the Radio Resource Control (RRC) protocol. The LPP positioning procedures usually consist of the following steps.

1) *Capability transfer:* The E-SMLC sends an *OTDOA-RequestCapabilities* message to the UE, and the UE responds with a *ProvideCapabilities* message to the E-SMLC. Example capabilities include supported frequency bands and inter-frequency RSTD measurements support.

2) *Assistance data transfer:* The E-SMLC sends a *ProvdieAssistanceData* message to the UE. This message contains the information of the specific PRS configuration of the suggested reference and neighbor cell list. With the assistance data, the UE knows when the PRS signals are transmitted and can measure TOA based on the PRS signals accordingly.

3) *Location information transfer:* The E-SMLC sends a *RequestLocationInformation* message to the UE, and the UE responds with a *ProvideLocationInformation* message to the E-SMLC within a certain response time.

Table 1: Comparison of Rel-9 LTE PRS, Rel-14 LTE-M PRS, and Rel-14 NPRS

| | LTE PRS (Rel-9) | LTE-M PRS (Rel-14) | NPRS (Rel-14) |
|---|---|---|---|
| Sequence generation | Pseudo-random QPSK sequence initialized by a parameter dependent on OFDM symbol index, slot index, and physical cell ID | Pseudo-random QPSK sequence initialized by a parameter dependent on OFDM symbol index, slot index, and PRS ID | Pseudo-random QPSK sequence initialized by a parameter dependent on OFDM symbol index, slot index, and PRS ID |
| Mapping to resource elements | • Frequency reuse 6 with physical cell ID dependent shift<br>• The center of the length-220 QPSK sequences are mapped to the resource elements in the center of the LTE carrier | • Frequency reuse 6 with PRS ID dependent shift<br>• The corresponding portion of the length-220 QPSK sequences are mapped to the resource elements in the specified location (which may not be the center of the LTE carrier) | • Frequency reuse 6 with PRS ID dependent shift<br>• For standalone or guardband, the central 2 elements of the length-220 QPSK sequences are mapped to the resource elements<br>• For inband, the corresponding portion of the length-220 QPSK sequences are mapped to the resource elements in the specified location |
| Transmission schedule | Periodic positioning occasions<br>• Periodicity: 160, 320, 640, or 1280 subframes<br>• Length of a positioning occasion: 1, 2, 4, or 6 subframes | Periodic positioning occasions<br>• Periodicity: 10, 20, 40, 80, 160, 320, 640, or 1280 subframes<br>• Length of a positioning occasion: 1, 2, 4, 6, 10, 20, 40, 80, or 160 subframes | Part A: a bitmap of length of 10 or 40 bits, with each bit indicating the presence of NPRS in the corresponding subframe<br>Part B: Periodic positioning occasions<br>• Periodicity: 160, 320, 640, or 1280 subframes<br>• Length of a positioning occasion: 10, 20, 40, 80, 160, 320, 640, or 1280 subframes |
| Frequency hopping | • No frequency hopping<br>• PRS fixed in the center of the LTE carrier | • Frequency hopping supported for PRS of 6 PRBs<br>• Can configure 2 or 4 different locations within the bandwidth of the wideband LTE carrier, with the first location fixed in the center of the LTE carrier<br>• PRS transmission cycles through the configured 2 or 4 frequency bands | • No explicit frequency hopping support<br>• PRS can be separately configured on multiple NB-IoT carriers, resulting in artificial frequency hopping |
| Muting | A bit string of 2, 4, 8, or 16 bits, with each bit applied to one positioning occasion | A bit string of 2, 4, 8, or 16 bits, with each bit applied to<br>• one positioning occasion, or<br>• one positioning occasion group consisting of 2, 4, 8, 16, 32, 64 or 128 positioning occasions | A bit string of 2, 4, 8, or 16 bits, with each bit applied to<br>• 10 consecutive subframes for configuration with Part A<br>• one positioning occasion for configuration with Part B |

The *ProvideLocationInformation* message includes the UE's RSTD measurement results.

Under the established LTE OTDOA architecture and positioning protocols, from a radio access perspective the design of PRS signals is the central topic to enhance OTDOA support of LTE-M or to enable OTDOA support of NB-IoT. LTE PRS is a well-established signal for TOA measurement in OTDOA since Release 9, and as will be explained later it offers 3 layers of isolation of PRS signals from different cells: 1) frequency domain with a reuse factor of 6, 2) time domain with muting, and 3) code domain with different cells using different sequences.

The designs of PRS for LTE-M and NB-IoT are built on LTE PRS. In the next two sections, we summarize the designs of PRS signals for LTE-M and NB-IoT separately, with a focus on the key aspects where the designs deviate from LTE PRS. A summary can be found in Table 1. Note that the PRS in NB-IoT is referred to as NPRS (a.k.a. Narrowband Positioning Reference Signal).

## IV. POSITIONING REFERENCE SIGNALS IN LTE-M

### A. PRS Sequence and Mapping to Resource Elements

The PRS sequence for LTE-M is the same as that of LTE. LTE PRS sequence is a pseudo-random Quadrature Phase Shift Keying (QPSK) sequence defined in [12]. The parameter used to initialize the pseudo-random sequence depends on the orthogonal frequency-division multiplexing (OFDM) symbol index, slot index, and physical cell identity. Accordingly, the

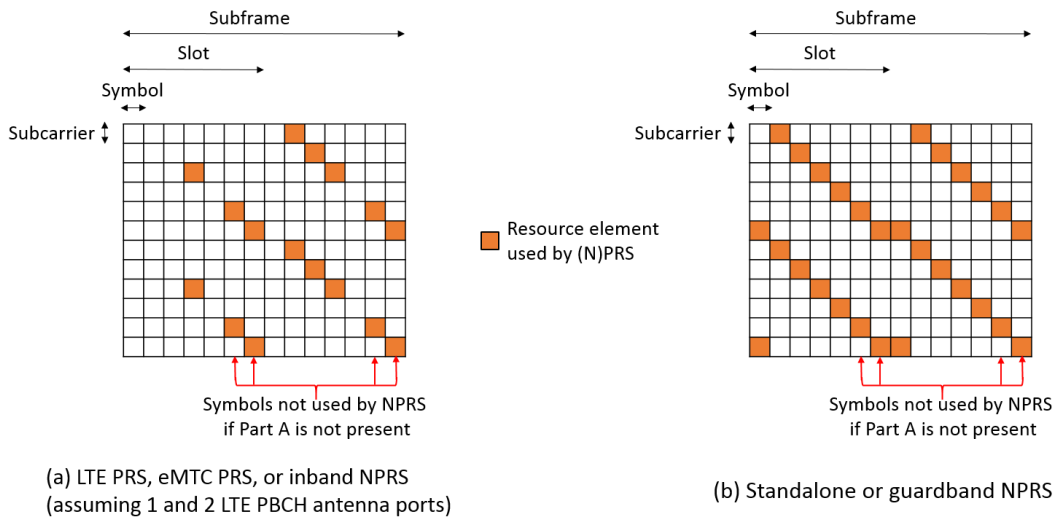

(a) LTE PRS, eMTC PRS, or inband NPRS
(assuming 1 and 2 LTE PBCH antenna ports)

(b) Standalone or guardband NPRS

Figure 2:Resource element mapping of positioning reference signals

sequence varies with the OFDM symbol index, slot index, and physical cell identity.

The overall time-frequency PRS mapping pattern is a diagonal pattern, as illustrated in Figure 2(a). Given an initialization parameter, a PRS sequence can be generated. The complex-valued QPSK elements of the sequence are then mapped to resource elements determined by physical cell identity. An important design feature of PRS is that PRS has a frequency reuse of 6. In LTE, each resource block consists of 12 subcarriers and the subcarrier spacing is 15 kHz. As is clear from Figure 2, if an OFDM symbol is used by PRS, only two of the 12 resource elements are used by PRS. By shifting the mapping in frequency, a total of 6 orthogonal PRS mappings are possible. The specific frequency shift is cell specific and is given by the physical cell identity mod 6.

Repeating the sequence generation and mapping, PRS sequences are generated as the time (i.e., symbol index and slot index) varies, with each sequence being mapped in frequency to the corresponding resource elements. Note that the first 3 OFDM symbols in a subframe are not used by PRS signals since they may be used by LTE L1/L2 signals such as Physical Downlink Control Channel (PDCCH). The other OFDM symbols not used by PRS in Figure 2 are used by LTE Cell-specific Reference Signal (CRS).

Note that LTE PRS signals are mapped to the central resource blocks of an LTE carrier, and the number of the LTE PRS resource blocks can be 6, 15, 25, 50, 75, or 100. For the bandwidth reduced and low complexity LTE-M UE, only 6-PRB wide signal can be received. To partially compensate the loss due to the reduced bandwidths of LTE-M UE, PRS frequency hopping is introduced. Each cell may configure 2 or 4 different locations within the bandwidth of the wideband LTE carrier, with the first location fixed in the center of the LTE carrier. Each frequency hopping band consists of 6 PRBs. The PRS transmission cycles through the configured 2 or 4 frequency bands, resulting in PRS frequency hopping.

### B. PRS Subframe Configuration

LTE PRS signals are transmitted in *positioning occasions*, with each positioning occasion consisting of 1, 2, 4, or 6 consecutive subframes. The positioning occasions occur periodically with a period $T_{PRS}$ of 160, 320, 640, or 1280 subframes. Each eNB can configure a subframe offset, which defines the starting subframe of PRS transmission relative to the start of a system frame cycle. The schedule of PRS transmission is illustrated in Figure 3.

With the existing LTE PRS, LTE-M UE can receive the PRS located in the central 6 resource blocks of an LTE carrier. However, this operation mode would limit the positioning performance of the wideband LTE UEs which can receive a wider PRS. It is important to ensure good performance for both device types with limited PRS resource overhead. To this end, a cell may configure and schedule multiple PRS transmissions, with each configuration and transmission schedule tailored to a device type. Then the LTE-M UE can receive longer PRS transmission to get the necessary number of PRS subframes to accurately measure the TOA, while the LTE UE receives the shorter wideband PRS transmission. This would reduce the PRS resource overhead in comparison to a single over-dimensioned PRS configuration with a wide bandwidth (for LTE UE) and a long positioning occasion (for LTE-M UE).

Note that LTE-M addresses coverage for deployments in challenging environments such as deep indoor scenario. Thus, it is necessary to introduce denser PRS transmissions compared with the existing LTE PRS configuration. This can be achieved by increasing the number of consecutive subframes per positioning occasion or reducing the positioning occasion periodicity. In LTE, as aforementioned, the maximum number of consecutive subframes in a positioning occasion is 6, and the minimum positioning occasion periodicity is 160 subframes. To support denser PRS transmission schedule, the set of numbers of subframes per positioning occasion has been extended to {1, 2, 4, 6, 10, 20, 40, 80, 160} subframes. Further, to support more frequent PRS transmissions in time, multiple positioning occasions may be configured in a legacy LTE PRS period. The positioning occasion interval in a PRS period can be 10, 20, 40, or 80 subframes.

Figure 3 gives an example of PRS configuration and transmission schedule, where 2 PRS transmissions are scheduled with the same subframe offset and period $T_{PRS}$. One

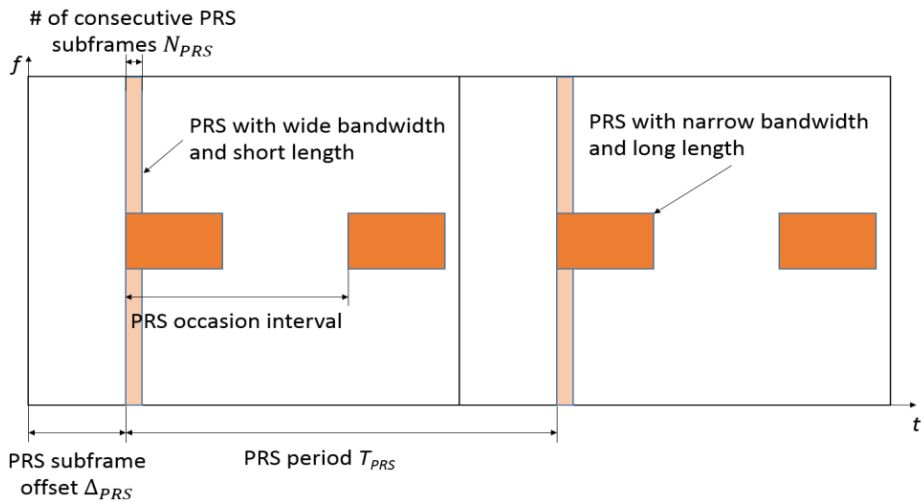

Figure 3: Positioning reference signals subframe configuration

PRS transmission with wide bandwidth and short PRS length is configured for LTE UE. Another PRS transmission with narrow bandwidth and long PRS length is configured for LTE-M UE. The second PRS transmission is more frequent with 2 positioning occasions per period. Thanks to the design synergies, the two PRS transmission schedules may overlap in time and frequency. This facilitates the sharing of PRS signals between LTE UE and LTE-M UE and optimizes radio resource usage.

### C. PRS Muting

To enable better detectability of the PRS from a weak cell that shares the same frequency shift with the PRS from a strong cell, PRS in the strong cell may be muted in certain occasions. This is known as PRS muting in LTE OTDOA. LTE PRS muting pattern is specified by a bit string of 2, 4, 8, or 16 bits for each cell. All the PRS subframes in one PRS positioning occasion are either all ON or all OFF as indicated by the corresponding bit in the bit string. The PRS muting in time helps achieve an effectively larger radio reuse factor beyond the frequency reuse of 6. The PRS muting helps mitigate interference at the cost of possibly increased time-to-fix of positioning. The optimal tradeoff between PRS hearability and time-to-fix of positioning depends on deployment scenarios, and it is up to the networks to decide on how to trade off the two metrics.

In LTE-M OTDOA, LTE PRS muting mechanism is largely reused. It should be however noted that, as described in Section IV-B, there may be multiple LTE-M PRS positioning occasions in a legacy LTE PRS period, and the interval between two positioning occasions of LTE-M PRS may be much shorter than that of LTE PRS. To facilitate the coordination of the muting of the PRS signals for different types of UEs, the maximum length of the bit string for LTE-M PRS signals would be 1024 bits (system frame cycle of 10240 subframes divided by the minimum positioning occasion interval of 10 subframes). Clearly, this would result in unacceptably large LPP assistance data overhead. The solution adopted is that each muting bit now applies to all the PRS positioning occasions in a legacy PRS period while the length of the bit string is still 2, 4, 8, or 16 bits.

In other words, each bit can indicate the presence of all the PRS subframes possibly belonging to multiple PRS positioning occasions in a PRS period. As a concrete example, with a bit string of two bits, "01", applied to the configuration in Figure 3, the 2 PRS positioning occasions of narrow bandwidth in every other PRS period are muted.

## V. NARROWBAND POSITIONING REFERENCE SIGNALS IN NB-IOT

### A. NPRS Subframe Configuration

For NB-IoT, NPRS is configured per NB-IoT carrier transmitting NPRS. Each NB-IoT carrier can have different configuration parameters, and there is no NPRS frequency hopping across NB-IoT carriers. The configuration may be indicated with two parts: Part A and Part B. The network can configure NPRS using Part A alone, or Part B alone, or both.

**Part A** uses a bitmap to indicate NPRS subframes in one NPRS positioning occasion. The value of each bit indicates the presence of NPRS in the corresponding subframe. The length of NPRS bitmap is either 10 bits or 40 bits, the same length as that of the bitmap of valid subframe configuration. The bitmap of valid subframe configuration was introduced in Release 13 to allow the networks to reserve some subframes for other purposes. The set of subframes that are indicated invalid are not used for transmission to NB-IoT UEs. Since Release-13 NB-IoT UEs do not understand NPRS transmission, the subframes containing NPRS shall be marked as invalid downlink subframes in the bitmap of valid subframe configuration.

Unlike the existing LTE PRS transmission schedule, the NPRS subframes indicated with Part A occur in every radio frame without any long-term periodicity. Equivalently, the length of the NPRS bitmap may be considered as the period of NPRS positioning occasions. Each NPRS period can be regarded as a positioning occasion, and the subframes used for NPRS are indicated by the NPRS bitmap. Clearly, with this indication, the NPRS subframes in a positioning occasion need not be consecutive, while the PRS subframes of a positioning occasion in LTE or LTE-M are always consecutive.

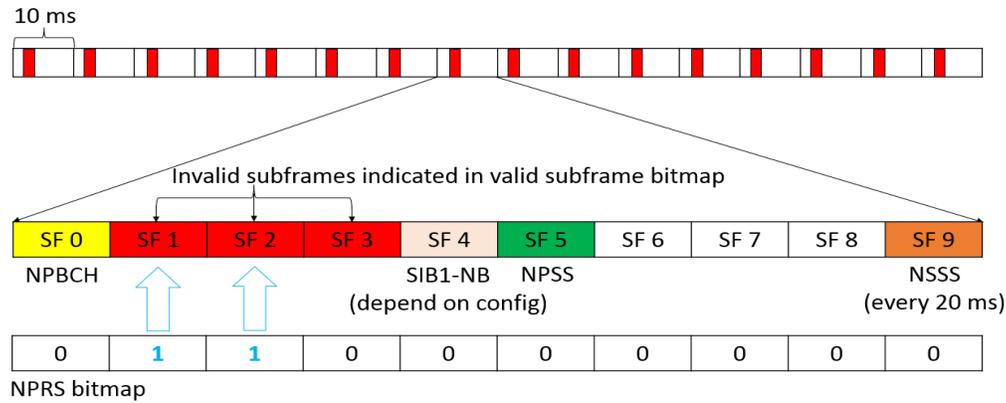

Figure 4: Narrowband positioning reference signals subframe configuration with Part A

Figure 4 illustrates an NPRS configuration indicated with Part A. In the example, the bitmap of valid subframe configuration is 10 bits long, and subframes 1, 2, and 3 are marked as invalid subframes. The NPRS bitmap is also 10 bits long and indicates that the invalid subframes 1 and 2 contain NPRS signals.

**Part B** is a configuration mechanism like that of LTE PRS. It specifies the periodicity of positioning occasions, the number of consecutive NPRS subframes in a positioning occasion, and the NPRS subframe offset. NPRS subframes indicated with Part B do not have to be invalid subframes. The periodicity of NPRS occasion is still chosen from the set {160, 320, 640, 1280} subframes. But the size of the set of NPRS subframe offset is limited to some extent to reduce the overhead of the LPP assistance data transfer. Specifically, for a given periodicity $T_{PRS}$ of positioning occasions, the NPRS subframe offset is $a*T_{PRS}$, where $a \in \{0, 1/8, 2/8, 3/8, 4/8, 5/8, 6/8, 7/8\}$. The number of consecutive NPRS subframes is chosen from the set {10, 20, 40, 80, 160, 320, 640, 1280} subframes. As a result, the consecutive NPRS transmission in a positioning occasion can be much longer than its counterparts in both LTE and LTE-M OTDOA. This long NPRS transmission is needed to partially compensate the reduced bandwidth of NPRS and to address coverage for deployments in challenging environments.

**Part A and Part B** may be both configured. In this case, a subframe contains NPRS if both parts of the configuration indicate that the subframe contains NPRS.

### B. NPRS Sequence and Mapping to Resource Elements

Like LTE PRS, NPRS resource element mapping is a diagonal pattern with a frequency reuse factor of 6. However, modifications have been introduced to cater for different deployment scenarios and/or different configuration parts.

For inband deployment where the NB-IoT carrier is deployed inside an LTE carrier (i.e., one LTE resource block is used by the NB-IoT carrier), NPRS resource element mapping depends on how NPRS subframes are configured.

- If NPRS subframes are configured with Part A or both Part A and Part B, NPRS signals are mapped to all OFDM symbols of an NPRS subframe except the first 3 OFDM symbols and the OFDM symbols that may be used by CRS. This NPRS mapping is the same as LTE PRS, as illustrated in Figure 2.

- If NPRS subframes are configured with Part B only, NPRS resource element mapping is the same as the mapping under Part A or both Part A and Part B, except that NPRS signals are not mapped to the last two OFDM symbols in each slot. This is because NPRS may be transmitted in valid NB-IoT downlink subframes that may contain narrowband reference signals (NRS). To avoid potential conflict between NPRS and NRS, NPRS signals are not mapped to the OFDM symbols that may be used by NRS. This NPRS resource element mapping is illustrated in Figure 2.

For standalone deployment (e.g., a GSM carrier of 200kHz is vacated for the NB-IoT carrier) or guardband deployment (e.g., the NB-IoT carrier is positioned in the guardband of a LTE carrier), NPRS signals can be mapped to more OFDM symbols than in inband deployment, since there is no need to avoid impacting on legacy LTE PDCCH region and LTE CRS that exist only in inband deployment. If NPRS subframes are configured with Part A or both Part A and Part B, NPRS signals are mapped to all OFDM symbols of an NPRS subframe, as illustrated in Figure 2. If NPRS subframes are configured with Part B only, NPRS signals are not mapped to the last two symbols in each slot, like the case of inband deployment.

As the NPRS resource element mapping pattern follows the diagonal pattern of LTE PRS in one resource block, NPRS sequence also reuses LTE PRS sequence, which as aforementioned is a pseudo-random QPSK sequence. The length of each LTE PRS sequence in frequency is 220. With two elements mapped onto one resource block, the LTE PRS sequence is long enough to be used in an LTE carrier whose maximum bandwidth is 100 resource blocks. Since the system bandwidth of a NB-IoT carrier is only 180 kHz, i.e., one resource block, a length-2 sequence is sufficient for NPRS. The design question is which 2 elements in the LTE PRS sequence should be used for NPRS.

For standalone or guardband deployment, the central two elements of LTE PRS sequence is used as NPRS sequence. Inband NB-IoT carriers can be flexibly placed in different positions inside a wideband LTE carrier. To maximize the synergy between LTE PRS and NPRS, the NPRS sequence is designed to be determined based on the position of the corresponding NB-IoT carrier inside the wideband LTE carrier. This also helps improve quasi-orthogonality of the pseudo-random sequences when NPRS signals are configured on the multiple inband NB-IoT carriers.

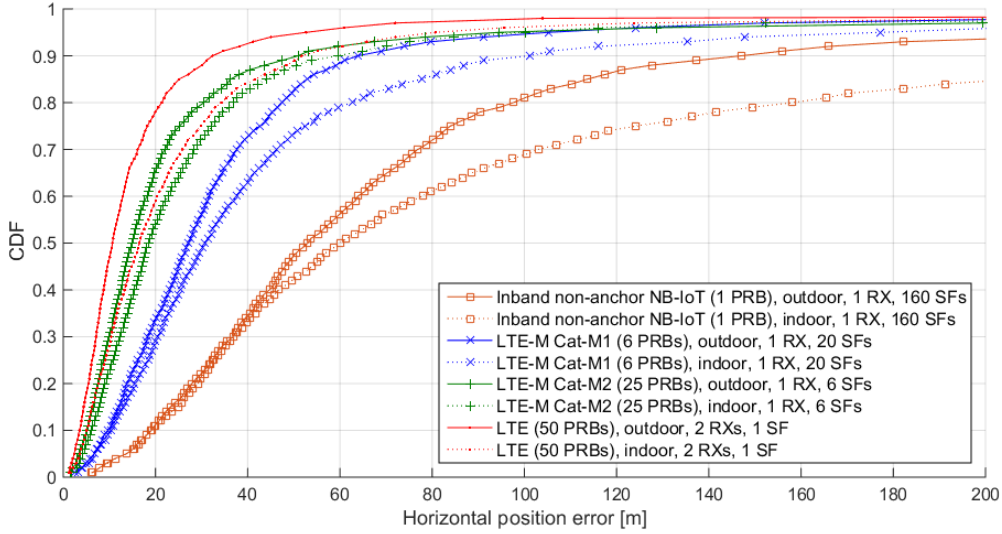

Figure 5: Positioning accuracy simulation of LTE, LTE-M, and NB-IoT: Macro only deployment with 700 m inter-site distance; perfect network synchronization; 700 MHz carrier frequency; 46 dBm eNB transmit power; 3 km/h UE speed; and 3D ITU channel model.

### C. NPRS Muting

Like LTE PRS muting, muting is also adopted for NPRS to enable better detectability of the PRS from a weak cell that shares the same frequency shift with the PRS from a strong cell. NPRS muting pattern is also specified by a bit string of 2, 4, 8, or 16 bits. On each NB-IoT carrier, one muting pattern may be signaled associated with Part A (if configured) and one muting pattern may be signaled associated with Part B (if configured). For Part A, a bit in a muting pattern indicates if the NPRS signals in consecutive 10 subframes are muted or not. For Part B, a bit in a muting pattern indicates if the NPRS signals in a positioning occasion are muted or not, which is the same as LTE PRS muting.

As aforementioned, if both Part A and Part B are used, a subframe contains NPRS if both parts of the configuration indicate that the subframe contains NPRS. As a result, with muting, an NPRS subframe is muted if it is muted by either Part A or Part B.

## VI. SIMULATION RESULTS AND DISCUSSIONS

In this section, we present some simulation results to gain insights into the OTDOA positioning performance by comparing Rel-9 LTE, Rel-14 LTE-M and NB-IoT. We focus on positioning accuracy and leave the other performance aspects such as energy consumption to future work. A deployment with a 10 MHz LTE carrier is assumed with the following configuration.

- LTE: 50-PRB wide PRS with a period of 160 ms and 1 PRS subframe in each positioning occasion.
- LTE-M Category M2: 25-PRB wide PRS with a period of 160 ms and 6 PRS subframes in each positioning occasion.

- LTE-M Category M1: 6-PRB wide PRS with a period of 160 ms and 20 PRS subframes in each positioning occasion.
- NB-IoT: NPRS is configured in every subframe on a non-anchor NB-IoT carrier in the LTE carrier.

The simulation results are shown in Figure 5, where 4-bit muting is used and each UE measures 8 PRS occasions per cell. It can be seen from Figure 5 that LTE provides the best positioning accuracy and that LTE-M outperforms NB-IoT under the simulation setup. This is expected since LTE has the largest signal bandwidth while NB-IoT has the smallest signal bandwidth. The performance loss due to bandwidth reduction cannot be fully compensated by longer PRS transmission. The positioning accuracy provided by OTDOA in LTE-M and NB-IoT can help enable LBS in a large set of IoT use cases.

Besides LTE-M and NB-IoT, there exist other low power wide area IoT connectivity technologies including Cooperative Ultra Narrowband (C-UNB) from Sigfox [13] and Long Range (LoRa) from Semtech [14] that use unlicensed spectrum. It is possible to develop and provide positioning support based on these non-3GPP technologies. For example, LoRa provides uplink based positioning, where the uplink signal from the device is detected by multiple LoRa gateways and the TOAs are estimated and processed to determine the position of the device [14]. Compared to OTDOA, the uplink based positioning may have lower device impact. But OTDOA has better scalability with the increasing number of IoT devices, because the configured PRS for OTDOA can be used by all the devices requiring positioning while in the uplink based positioning each device needs to transmit an uplink signal.

## VII. CONCLUSIONS AND FUTURE WORK

Positioning is vital for many IoT applications. In this article, we provide an overview of the 3GPP's work in enhancing the positioning support for both LTE-M and NB-IoT during its Release 14, with a focus on OTDOA. The enhanced positioning support will further improve the 3GPP-based IoT eco-system and increase the market impact of 3GPP IoT technologies.

Further evolution of the positioning support for LTE-M and NB-IoT can be envisioned. One possible direction is to enhance the support of uplink based positioning methods based on e.g. sounding reference signals for LTE-M or narrowband physical random access signals for NB-IoT [15]. Another interesting direction is to investigate integrated OTDOA positioning for advanced UE that can receive and process all the positioning reference signals configured by the network. Also, massive LTE-M and NB-IoT devices are expected to have a long life cycle that may go beyond network evolution toward 5G. It will be beneficial that future 5G positioning would be designed to have some synergies with the LTE-M and/or NB-IoT positioning.

## REFERENCES

[1] H. Shariatmadari, R. Ratasuk, S. Iraji, A. Laya, T. Taleb, R. Jäntti, and A. Ghosh, "Machine-type communications: current status and future perspectives toward 5G systems." *IEEE Communications Magazine*, vol. 53, no. 9, pp. 10-17, September 2015.

[2] 3GPP TS 22.368, "Service requirements for machine-type communications (MTC)," V13.1.0, December 2014.

[3] 3GPP TS 22.261, "Service requirements for next generation new services and markets," V1.0.0, December 2016.

[4] A. Rico-Alvarino, M. Vajapeyam, H. Xu, X. Wang, Y. Blankenship, J. Bergman, T. Tirronen, and E. Yavuz, "An overview of 3GPP enhancements on machine to machine communications," *IEEE Communications Magazine*, vol. 54, no. 6, pp. 14-21, June 2016.

[5] Y.-P. E. Wang, X. Lin, A. Adhikary, A. Grövlen, Y. Sui, Y. Blankenship, J. Bergman, and H. S. Razaghi, "A primer on 3GPP narrowband internet of things," *IEEE Communications Magazine*, vol. 53, no. 3, pp. 117-123, March 2017.

[6] 3GPP R1-167101, "Discussion on positioning requirements for enhanced NB-IoT," CMCC, August 2016.

[7] 3GPP R1-167743, "Requirements for NB IoT positioning enhancements evaluations," Vodafone, August 2016.

[8] 3GPP R1-167790, "Requirements for NB-IoT/eMTC positioning," Deutsche Telekom, August 2016.

[9] 3GPP TS 36.305, "Evolved Universal Terrestrial Radio Access Network (E-UTRAN); Stage 2 functional specification of User Equipment (UE) positioning in E-UTRAN," V14.0.0, December 2016.

[10] S. Fischer, "Observed Time Difference Of Arrival (OTDOA) Positioning in 3GPP LTE," white paper, Qualcomm Inc., June 2014. Available at https://www.qualcomm.com/media/ documents/files/otdoa-positioning-in-3gpp-lte.pdf

[11] 3GPP TS 36.355, "Evolved Universal Terrestrial Radio Access (E-UTRA); LTE Positioning Protocol (LPP)," V14.0.0, December 2016.

[12] 3GPP TS 36.211, "Evolved Universal Terrestrial Radio Access (E-UTRA); Physical channels and modulation," V14.1.0, January 2017.

[13] 3GPP GP-150057, "C-UNB technology for cellular IoT - physical layer," Sigfox Wireless, March 2015.

[14] 3GPP GP-150074, "Combined narrow-band and spread spectrum physical layer proposal for cellular IoT," Semtech, March 2015.

[15] X. Lin, A. Adhikary, and Y.-P. E. Wang, "Random access preamble design and detection for 3GPP narrowband IoT systems," *IEEE Wireless Communications Letters*, vol. 5, no. 6, pp. 640-643, December 2016.